\documentstyle[aps]{revtex}
\input epsf
\input psfig
\begin{document}

\draft
\title{How to fool CMB parameter estimation}
\author{William H.\ Kinney\thanks{Electronic address: {\tt 
kinney@phys.ufl.edu}}}
\address{Dept. of Physics, University of Florida}
\address{P.O. Box 118440, Gainesville, FL 32611}
\date{\today}
\maketitle

\begin{abstract}
With the release of the data from the Boomerang and MAXIMA-1 balloon flights, 
estimates of cosmological parameters based on the Cosmic Microwave Background 
(CMB) have reached unprecedented precision. In this paper I show that it is 
possible for these estimates to be substantially biased by features in the 
primordial density power spectrum. I construct primordial power spectra which 
mimic to within cosmic variance errors the effect of changing parameters such as 
the baryon density and neutrino mass, meaning that even an ideal measurement 
would be unable to resolve the degeneracy. Complementary measurements are 
necessary to resolve this ambiguity in parameter estimation efforts based on CMB 
temperature fluctuations alone.
\end{abstract}


\section{Introduction}
\label{secintro}

With the release of the Boomerang\cite{boom2000} and MAXIMA-1\cite{MAXIMAa} data 
sets, the promise of using the Cosmic Microwave Background to provide precision 
constraints on cosmological parameters has become a reality. The inflationary 
prediction of a flat universe has received spectacular confirmation, and 
remarkably good estimates of other cosmological parameters have 
emerged\cite{boom2000a,MAXIMAb}. More detailed observations of the CMB, in 
particular NASA's MAP satellite\cite{MAP} and the ESA's Planck 
Surveyor\cite{Planck}, should allow for exquisitely precise determination of a 
handful of  ``fundamental'' cosmological parameters such as the baryon density 
and the Hubble constant. 

However, current parameter estimation efforts based on observation of the CMB 
suffer from an inherent ambiguity: lack of knowledge about the primordial 
density fluctuation spectrum. Primordial density and gravity wave fluctuations 
are the underlying source of the fluctuations in the CMB. The ability to predict 
the form of the CMB temperature anisotropy depends on knowledge of the form of 
these fluctuations. The standard assumption is that the density fluctuation 
power spectrum is a featureless power law, $P\left(k\right) \propto k^{n}$. Such 
an assumption is theoretically supported by inflation, which predicts such a 
power law spectrum. It is also empirically supported by observation of large 
scale structure. Current large scale structure data, however, only loosely 
constrain the form of the primordial power spectrum on scales relevant for CMB 
anisotropy measurements, $k \leq 0.1\, h\,{\rm MpC}^{-1}$\cite{gawiser98}. A 
fully empirical approach to the data forces one to consider an {\it arbitrary} 
density power spectrum as input to the parameter estimation problem. This in 
effect adds so many free parameters (and associated degeneracies) that precision 
determination of parameters from the CMB temperature anisotropy alone becomes an 
impossibility. In this paper I expicitly construct primordial power spectra that 
mimic the effects of changes in cosmological parameters to within cosmic 
variance errors, meaning that even a perfect measurement of the CMB temperature 
fluctuations would be unable to resolve the degeneracy. This requires large 
deviations from the standard scale-invariant power law spectrum, $\delta P(k) / 
P(k) \sim 0.5$, which would be detectable with complementary measurements, 
such as CMB polarization and improved large-scale structure data like that from 
the Sloan Digital Sky Survey\cite{sloandss}. 

\section{The primordial power spectrum and the CMB}
\label{CMBbackground}

All observations of the CMB anisotropy published to date have been maps of the 
anisotropy in the temperature of the CMB, $\Delta T / T_0 \sim 10^{-5}$, where 
$T_0 = 2.728\,{\rm K}$. It is convenient to quantify these temperature 
fluctuations as a set of multipole moments,
\begin{equation}
{\delta T\left(\theta,\phi\right) \over T_0} = \sum_{\ell = 0}^{\infty}\sum_{m = 
-\ell}^{\ell}{a_{\ell m} Y_{\ell m}\left(\theta,\phi\right)}.
\end{equation}
Lack of a preferred direction implies that the amplitudes $a_{\ell m}$ will be 
independent of $m$, 
\begin{equation}
\left\langle a^{*}_{\ell' m'} a_{\ell m}\right\rangle = C_{\ell} \delta_{\ell 
\ell'} \delta_{m m'},
\end{equation}
where the angle brackets denote an average over realizations. For Gaussian 
fluctuations, the set of $C_{\ell}$'s completely characterizes the temperature 
anisotropy. If the fluctuations are non-Gaussian, higher order correlation 
functions are necessary to fully characterize the anisotropy.  The power of the 
CMB in constraining cosmological parameters comes from the fact that one is 
using a large number of parameters (the $C_\ell$ spectrum) to constrain a dozen 
or so ``fundamental'' parameters such as the baryon density, the ratio of 
gravity wave perturbations to density perturbations, and so on. Any given CMB 
observation will have access to a range of $C_{\ell}$'s with error bars 
determined by a combination of the observation's sky coverage, sensitivity and 
angular resolution. However, there exists a fundamental limit to the accuracy 
with which the $C_{\ell}$ spectrum can be measured, referred to as {\it cosmic 
variance},
\begin{equation}
{\Delta C_\ell \over C_\ell} \geq \sqrt{1 \over 2 \ell + 1}.
\end{equation} 
Cosmic variance is simply a finite sample size effect coming from the fact that 
there is only a single sky to measure, and is more significant at long 
wavelength (smaller $\ell$). Cosmic variance will play a central role in the 
discussion below for the reason that any two $C_{\ell}$ spectra which have a 
small enough $\chi^2$ over cosmic variance errors are {\it in principle} 
indistinguishable, even when subject to an ideal measurement. 

Going from a $C_{\ell}$ spectrum to cosmological parameters requires assumptions 
about the form of the density power spectrum. A typical assumption 
when discussing the ability to constrain cosmological parameters using the CMB 
is that the primordial density power spectrum is a power law:
\begin{equation}
P\left(k\right) \propto k^n,
\end{equation}
where the spectral index $n$ is close to the scale-invariant value $n = 1$. The 
recent Boomerang measurement\cite{boom2000a} yields a value $n = 0.87^{+ 
0.11}_{-0.08}$ for a set of priors consistent with inflation, and the MAXIMA-1 
measurement has a best fit of $n = 0.99 \pm 0.09$\cite{MAXIMAb}. Such a 
featureless power law spectrum is in fact predicted by most inflationary models. 
A natural extension of this approximation is to allow ``running'' of the 
spectral index,
\begin{equation}
{d n \over d \log k} \neq 0.
\end{equation}
Such running is a feature of some inflation 
models\cite{stewart97,stewart97a,copeland97,covi98,covi99,covi00}. Future CMB 
observations will be able to detect running of the spectral index as small as $d 
n / d \log\left(k\right) \simeq 0.05$\cite{copeland97}. Taking this a step 
further, Lesgourges, Prunet and Polarski considered CMB constraints on models 
with broken scale invariance\cite{lesgourgues99}. In principle, however, the 
primordial power spectrum can contain many more free parameters than just one or 
two. The purpose of this paper is to investigate the possibility that features 
in the primordial power spectrum could mimic the effect of other parameters and 
thus ``fool'' parameter estimation efforts which assume a featureless power law 
spectrum. It is to be expected that such confusion will be possible, since the 
number of parameters available in an arbitrary primordial power spectrum is much 
larger than the parameter space considered by typical parameter estimation 
efforts, and can in principle be larger than the number of multipoles available 
in the CMB for measurement. I show below that the number of free 
parameters necessary to confuse parameter estimation efforts is fewer than a 
hundred. In the next section I discuss the numerical methods used to approach 
the problem.

\section{Numerical methods}
\label{numericalmethods}

Boltzmann codes for calculating the CMB spectrum normally take a set of 
cosmological parameters as input and generate a spectrum of $C_{\ell}$'s. In 
this paper I construct primordial power spectra which would cause significant 
mis-estimation of parameters if a power law fluctuation spectrum were to be 
assumed in the likelihood analysis. This amounts to solving the inverse problem 
-- given a particular $C_{\ell}$ spectrum, what primordial power spectrum is 
necessary to produce it? I adopt a simple $\chi^2$ minimization procedure for 
solving this problem. The procedure is as follows.

I choose two models and generate CMB multipole spectra for each using the 
CMBFAST code\cite{seljak97}. One I call the ``true'' model, with a set of 
parameters assumed to represent the actual underlying cosmos. The second I call 
the ``target'' model, generated with a different set of parameters but the {\it 
same density power spectrum}, which I take to be the scale invariant spectrum 
$P\left(k\right) \propto k$. The goal is to deform the density power spectrum in 
such a way as to mimic the $C_{\ell}$ spectrum of the target model while 
retaining the underlying parameters of the true model. The observer will see the 
target spectrum, but the true parameters of the universe are those of the true 
model. The input power spectrum is binned into $N_k = 75$ bins $i$ with 
wavenumber $k_i$. This binning provides a good balance between accuracy and 
computational efficiency. CMBFAST samples the power spectrum at a 
much higher resolution, so the binned data is smoothed using cubic spline 
interpolation\cite{nrcubicspline}. The ``trial'' density power spectrum $P^{\rm 
trial}\left(k\right)$ is initially a power law and is adjusted iteratively until 
the trial CMB spectrum $C^{\rm trial}_{\ell}$ matches the target CMB spectrum 
$C^{\rm target}_{\ell}$ to within cosmic variance errors. The first step is to 
numerically calculate a correlation matrix between the $C_{\ell}$'s and the 
primordial power spectrum
\begin{equation}
D_{i \ell} \equiv {\partial C^{\rm trial}_{\ell} \over \partial 
P\left(k_i\right)},\label{eqderivativematrix}
\end{equation}
where $k_i$ is the wavenumber of the $i$'th bin and $P\left(k_i\right)$ is the 
amplitude of the density power spectrum at that scale.\footnote{The structure of 
the CMBFAST software is particularly well suited to this calculation, as the 
Boltzmann integration needs to be run only once to calculate the entire 
derivative matrix.} Note that this is not 
a square matrix: $\ell = 1, \ldots N_\ell$ and $i = 1, \ldots ,N_k$. I use $N_k 
= 75$ and $N_\ell = 1500$. It is also necessary to compute the derivative of 
the $\chi^2$ between the trial and target spectra:
\begin{eqnarray}
{\partial \chi^2 \over \partial P\left(k_i\right)} =&& {\partial \over \partial 
P\left(k_i\right)} \sum_{\ell}^{}{ \left[{C^{\rm trial}_\ell - C^{\rm 
target}_\ell \over \sigma_\ell}\right]^2}\cr
=&& 2 \sum_{\ell}{D_{i \ell} {\left[C^{\rm trial}_\ell - C^{\rm 
target}_\ell\right] \over 
\sigma_l^2} }.
\end{eqnarray}
I take the errors to be the cosmic variance limit
\begin{equation}
\sigma_\ell = \sqrt{1 \over 2 \ell + 1} C^{\rm target}_\ell.
\end{equation}
This is particularly important: any two spectra that agree within cosmic 
variance limits are impossible to distinguish with {\it any} CMB measurement, no 
matter how sensitive. The result is therefore independent of any particular 
experiment. A new density power spectrum is calculated by gradient 
descent\cite{nrgradientdescent}:
\begin{equation}
P^{\rm new}\left(k_i\right) = P^{\rm trial}\left(k_i\right) - \lambda \sum_{j = 
1}^{N_k}{\alpha^{-1}_{i j} {\partial \chi^2 \over \partial P\left(k_j\right)}},
\end{equation}
where $\lambda$ is a constant whose value is picked to ensure smooth convergence 
(I use $\lambda = 0.1$), and the matrix $\alpha_{i j}$ is defined in terms of 
the derivative matrix (\ref{eqderivativematrix}) as:
\begin{equation}
\alpha_{i j} \equiv \sum_{\ell = 1}^{N_\ell}{{1 \over \sigma_l^2} D_{i \ell} 
D_{j \ell}}.
\end{equation}
Inversion of the matrix $\alpha_{i j}$ is in practice somewhat problematic, as 
the matrix is in numerically singular. Typically only half of the 75 
eingenvalues of the matrix are finite. I use singular value decomposition 
methods to perform the inversion\cite{nrsvdcomp}. Once a new $P^{\rm trial}$ is 
generated, the procedure is iterated until an acceptable $\chi^2$ is achieved, 
which I define to be $\chi^2 < 100$, more than good enough to make the spectra 
indistinguishable in practice. Convergence typically requires around 20 
iterations. In the next section I describe the results of the calculations.

\section{Results and Conclusions}
\label{secresults}

I choose a target model that is a good fit to the Boomerang and MAXIMA-1 
observations, with the following parameters:
\begin{eqnarray}
&&\Omega_{\rm total} = 1\cr
&&\Omega_{\rm b} h^2 = 0.027 \cr 
&&\Omega_{CDM} = 0.3\cr 
&&\Omega_{\Lambda} = 0.645 \cr 
&&\Omega_{\nu} = 0\cr 
&&h = 0.7\cr 
&&n = 1\cr
&&\tau = 0.2\cr
&&r = 0
\end{eqnarray}
Here $h$ is the Hubble constant in units of $100\,{\rm km/s}$, $n$ is the scalar 
spectral index, $\tau$ is the reionization optical depth, and $r$ is the 
tensor/scalar ratio. The procedure is to change one or more of the parameters of 
the target model to produce a ``true'' model, then reproduce the $C_{\ell}$ 
spectrum of the target model by modifying the underlying matter power spectrum 
as described in Section \ref{numericalmethods}. This choice of target model has 
two peculiarities. First, the baryonic density is outside the range preferred by 
Big Bang Nucleosynthesis (BBN), $\Omega_{\rm b} h^2 = 0.019 \pm 
0.0024$\cite{burles98,burles99}. Second, the reionization optical depth is 
large, $\tau = 0.2$. Figure 1 shows the target $C_{\ell}$ spectrum and the 
$C_{\ell}$ spectrum resulting from setting $\Omega_{\rm b} h^2 = 0.019$ in 
accordance with BBN limits. The power spectrum necessary to change the 
$C_{\ell}$ spectrum of the true model into one indistinguishable from that of 
the target model is shown in Figure 2. The perturbation to the power spectrum is 
quite large, and would be difficult to produce via inflation. However, such a 
power spectrum is in principle allowed by existing data. Another choice is to 
change the reionization optical depth $\tau$ instead of the baryon density. Fig. 
3 shows the target model and a model with $\tau = 0$. The power spectrum 
required for the true model to mimic the target model is shown in Figure 4. In 
this case, the power spectrum is dominated by a single large feature at long 
wavelength, a situation that could potentially be realized within an 
inflationary context. Finally, Figures 5 and 6 show the results with $\Omega_\nu 
= 0.1$, corresponding to a neutrino mass of $m_\nu = 4.5\,{\rm eV}$. 

It is clear that allowing for an arbitrary density power spectrum renders 
attempts at parameter estimation from the CMB temperature anisotropy alone a 
hopeless task. It is necessary to use complementary measurements to constrain 
the density power spectrum.  If measurements of the CMB polarization are 
available in addition to measurements of the temperature anisotropy, the task of 
``fooling'' CMB parameter estimation efforts becomes much more difficult. 
Polarization is a tensor quantity, which can be decomposed on the celestial 
sphere into ``electric-type'', or scalar, and ``magnetic-type'', or pseudoscalar 
modes. The symmetric, trace-free polarization tensor ${\cal P}_{ab}$ can be 
expanded as\cite{kamionkowski96}
\begin{equation}
{{\cal P}_{ab} \over T_0} =  \sum_{\ell = 0}^{\infty}\sum_{m = -\ell}^{\ell} 
\left[a^E_{\ell m} Y^E_{\left(\ell m\right) ab}\left(\theta,\phi\right) + 
a^B_{\ell m} Y^B_{\left(\ell m\right) ab}\left(\theta,\phi\right)\right],
\end{equation}
where the $Y^{E,B}_{\left(\ell m\right) a b}$ are electric- and magnetic-type 
tensor spherical harmonics, with parity $(-1)^\ell$ and $(-1)^{\ell + 1}$, 
respectively. Unlike a temperature-only map, which is described by the single 
multipole spectrum of $C^T_\ell$'s, a temperature/polarization map is described 
by three spectra
\begin{equation}
\left\langle \left|a^T_{\ell m}\right|^2\right\rangle \equiv C_{T\ell},\ 
\left\langle\left|a^E_{\ell m}\right|^2\right\rangle \equiv C_{E\ell},\ 
\left\langle\left|a^B_{\ell m}\right|^2\right\rangle \equiv C_{B\ell},
\end{equation}
and three correlation functions,
\begin{equation}
\left\langle a^{T*}_{\ell m} a^E_{\ell m}\right\rangle \equiv C_{C\ell},\ 
\left\langle a^{T*}_{\ell m} a^B_{\ell m} \right\rangle \equiv C_{(TB)\ell},\ 
\left\langle a^{E*}_{\ell m} a^B_{\ell m}\right\rangle \equiv C_{(EB)\ell}.
\end{equation}
Parity requires that the last two correlation functions vanish, $C_{(TB)\ell} = 
C_{(EB)\ell} = 0$, leaving four spectra: temperature $C_{T\ell}$, E-mode 
$C_{E\ell}$, B-mode $C_{B\ell}$, and the cross-correlation $C_{C\ell}$. Since 
scalar density perturbations have no ``handedness,'' it is impossible for scalar 
modes to produce B-mode (pseudoscalar) polarization. Only tensor fluctuations 
(or foregrounds \cite{zaldarriaga98}) can produce a B-mode. In the cases I have 
considered in this paper, calculation of the polarization spectra reveals that 
models with indistinguishable temperature fluctuations have very distinct 
patterns of polarization, so that a sufficiently sensitive measurement of the 
polarization power spectra (such as that which will be produced by the Planck 
satellite) would reveal any inconsistency in the assumption of a 
power-law density fluctuation spectrum. Figure 7 shows the E-mode polarization 
spectra for the case considered in Fig. 1, in which the baryon density is 
varied. Fig. 8 shows the cross-correlation spectrum $C_{T\ell}$ for the same 
case. The B-mode vanishes. Table I shows the $\chi^2$ over cosmic variance 
errors for all three CMB multipole spectra (temperature, E-mode and 
cross-correlation) for the models considered in this paper. In all three cases, 
although it is possible to mimic the $C_{T \ell}$ spectrum of the target model 
to within cosmic variance, the polarization spectra produced are poor fits to 
the respective target polarization spectra. Additional information on the 
density power spectrum can be obtained from large scale structure data. The 
prostpect for independent constraint of the density power spectrum from a 
combination of SDSS data and CMB polarization was considered by Wang {\it et 
al.}\cite{wang98}.

\vbox{
\begin{table}
\begin{center}
\begin{tabular}{c|c|c|c}
Model &
$\chi^2$: Temperature & $\chi^2$: E-mode & $\chi^2$: cross-correlation\cr
\hline \hline
$\Omega_{\rm b} h^2 = 0.019$,\ $\Omega_{\Lambda} = 0.661$ & $80$ & $2.2 \times 
10^{5}$ & $4.9 \times 10^7$\cr
\hline
$\tau = 0$ & $76$ & $500$ & $2.1 \times 10^4$\cr
\hline
$\Omega_\nu = 0.1$,\ $\Omega_{\Lambda} = 0.545$ &  $97$  &  $3.5 \times 10^4$ & 
$1.8 \times 10^{7}$
\end{tabular}
\bigskip
\bigskip
\caption{$\chi^2$ for various model fits, relative to cosmic variance errors 
over $N_\ell = 1500$ multipole moments. All $\chi^2$ values are calculated 
relative to the target model, with $\Omega_{\rm b} h^2 = 0.027$, $\Omega_\Lambda 
= 0.645$, and $\tau = 0.2$. $\Omega_{\rm total} = 1$ in all cases. While the 
temperature multipole spectra are excellent fits, the fits of the polarization 
spectra are poor. }
\end{center}
\end{table}}

In this paper I have shown by construction that it is possible to mimic the 
effect of changes in fundamental cosmological parameters on the CMB by changes 
in the density power spectrum. With a $N_k = 75$ parameter fit of the power 
spectrum, a CMB multipole spectrum with significantly shifted parameters can be 
fit to within cosmic variance errors, so that even an ideal measurement would be 
unable to resolve the degeneracy. It is therefore possible for estimates of 
cosmological parameters based on CMB temperature fluctuations alone to be 
substantially biased by features in the density power spectrum. This degeneracy 
cannot be resolved without access to complementary measurements such as 
observation of CMB polarization or information about large scale structure such 
as that which will be provided by the Sloan Digital Sky Survey.

\section*{Acknowledgments}

This work was supported in part by U.S. DOE and NASA grant NAG5-7092 at Fermilab 
and U.S. DOE grant DE-FG02-97ER-41029 at University of Florida. I would like to 
thank Edward Kolb for invaluable discussions.

\vfill\eject

\begin{figure}
\label{figomegabCl}
\psfig{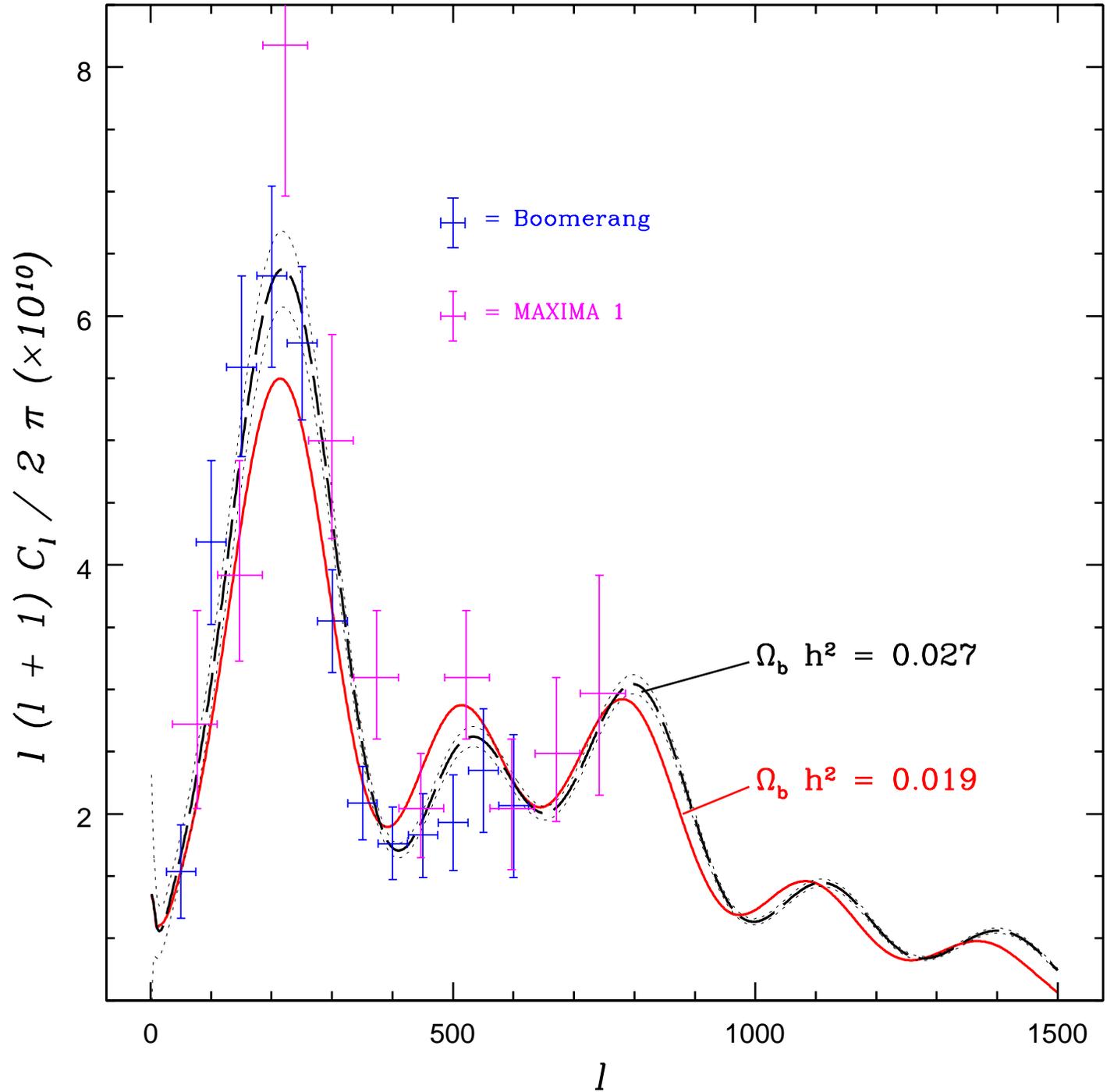}
\bigskip
\bigskip
\bigskip
\bigskip
\bigskip
\bigskip
\caption{$C_{\ell}$ spectra for target model (black, dashed) and model with 
$\Omega_{\rm b} h^2 = 0.019$ (red, solid). The spectra differ by $\chi^2 = 1.4 
\times 10^5$ over 1500 degrees of freedom, with cosmic variance errors. The 
dotted lines show the cosmic variance ``envelope'' for the target model. Error 
bars are the Boomerang and MAXIMA-1 measurements, shown for comparison purposes 
only.}
\end{figure}

\begin{figure}
\psfig{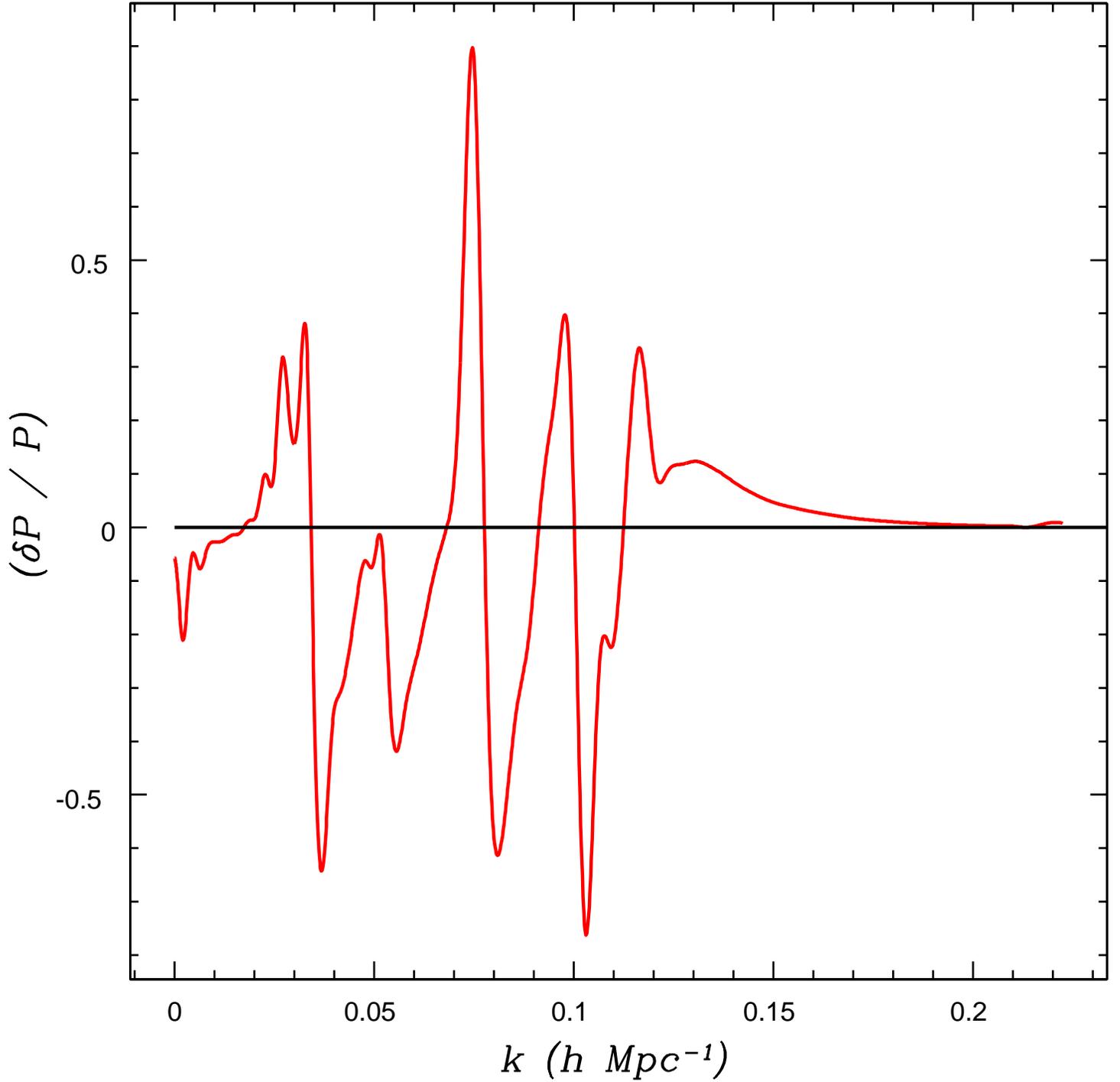}
\bigskip
\bigskip
\bigskip
\bigskip
\bigskip
\bigskip
\caption{Density power spectrum which mimics the target model ($\Omega_{\rm b} 
h^2 = 0.027$) with parameters from the true model, $\Omega_{\rm b} h^2 = 0.019$. 
With respect to cosmic variance errors, the fit has $\chi^2 = 81$ over 
$1500$ degrees of freedom in the $C_{\ell}$ spectrum. On this plot $\delta P / P 
\equiv P(k) / k - 1$.}
\end{figure}

\begin{figure}
\psfig{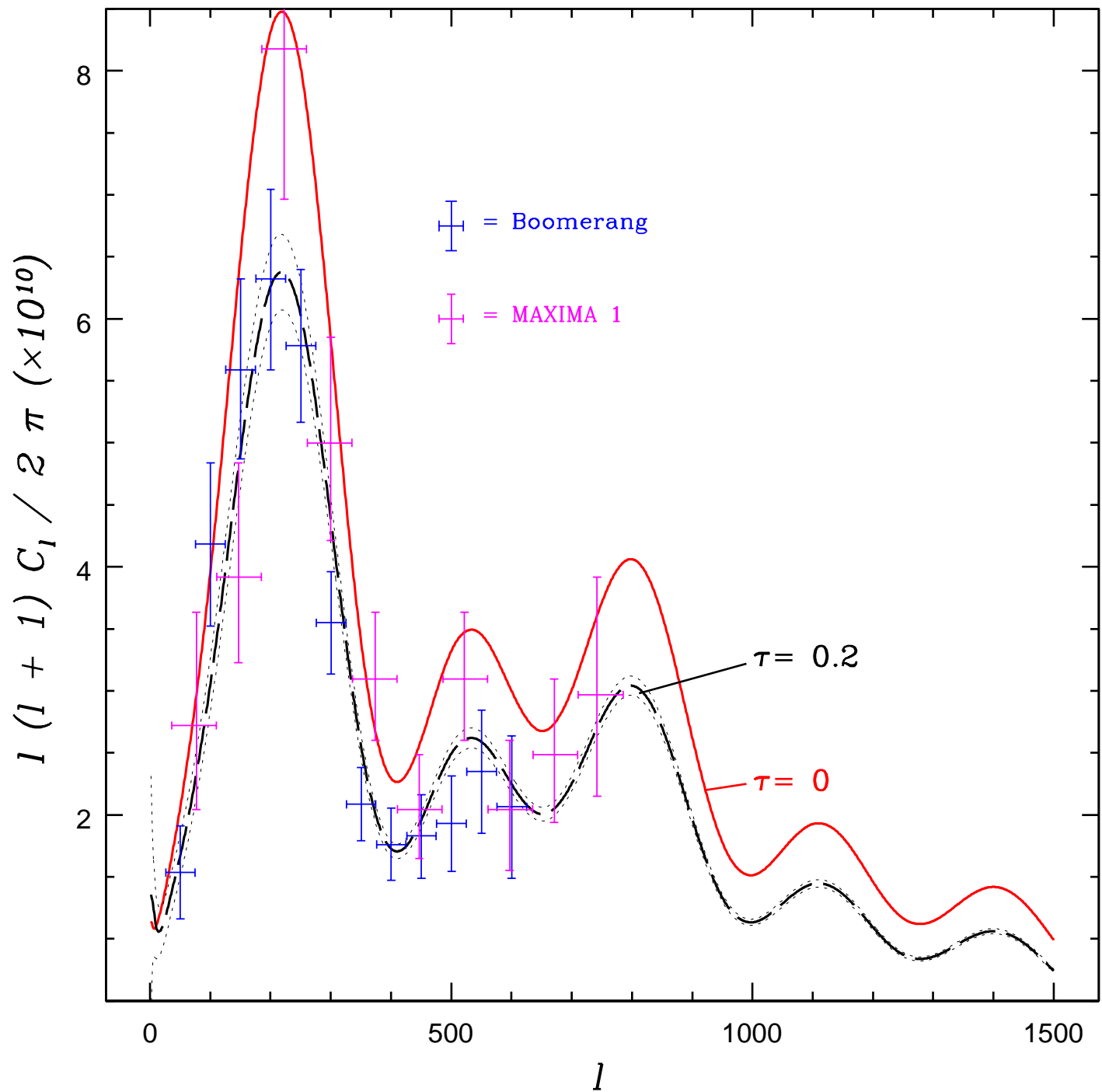}
\bigskip
\bigskip
\bigskip
\bigskip
\bigskip
\bigskip
\caption{$C_{\ell}$ spectra for the target model (black, dashed) and a true 
model with $\tau = 0$ (red, solid). The spectra differ by $\chi^2 = 1.3 \times 
10^5$ over 1500 degrees of freedom.}
\end{figure}

\begin{figure}
\psfig{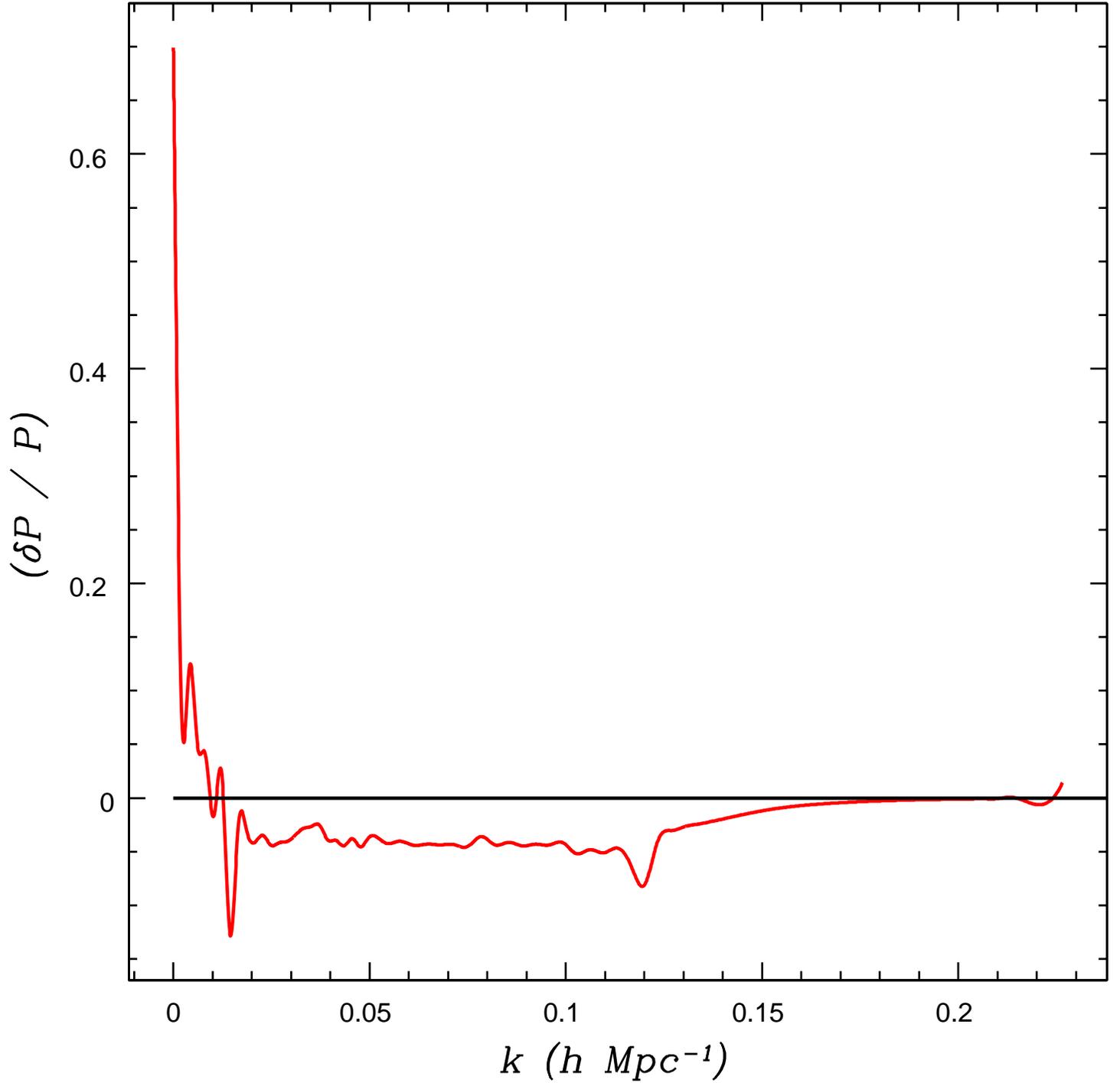}
\bigskip
\bigskip
\bigskip
\bigskip
\bigskip
\bigskip
\caption{Density power spectrum which mimics the $C_{\ell}'s$ of the target 
model ($\tau = 0.2$) with parameters from the true model, $\tau = 0$. The final 
fit has  $\chi^2 = 76$ over 1500 degrees of freedom.}
\end{figure}

\begin{figure}
\psfig{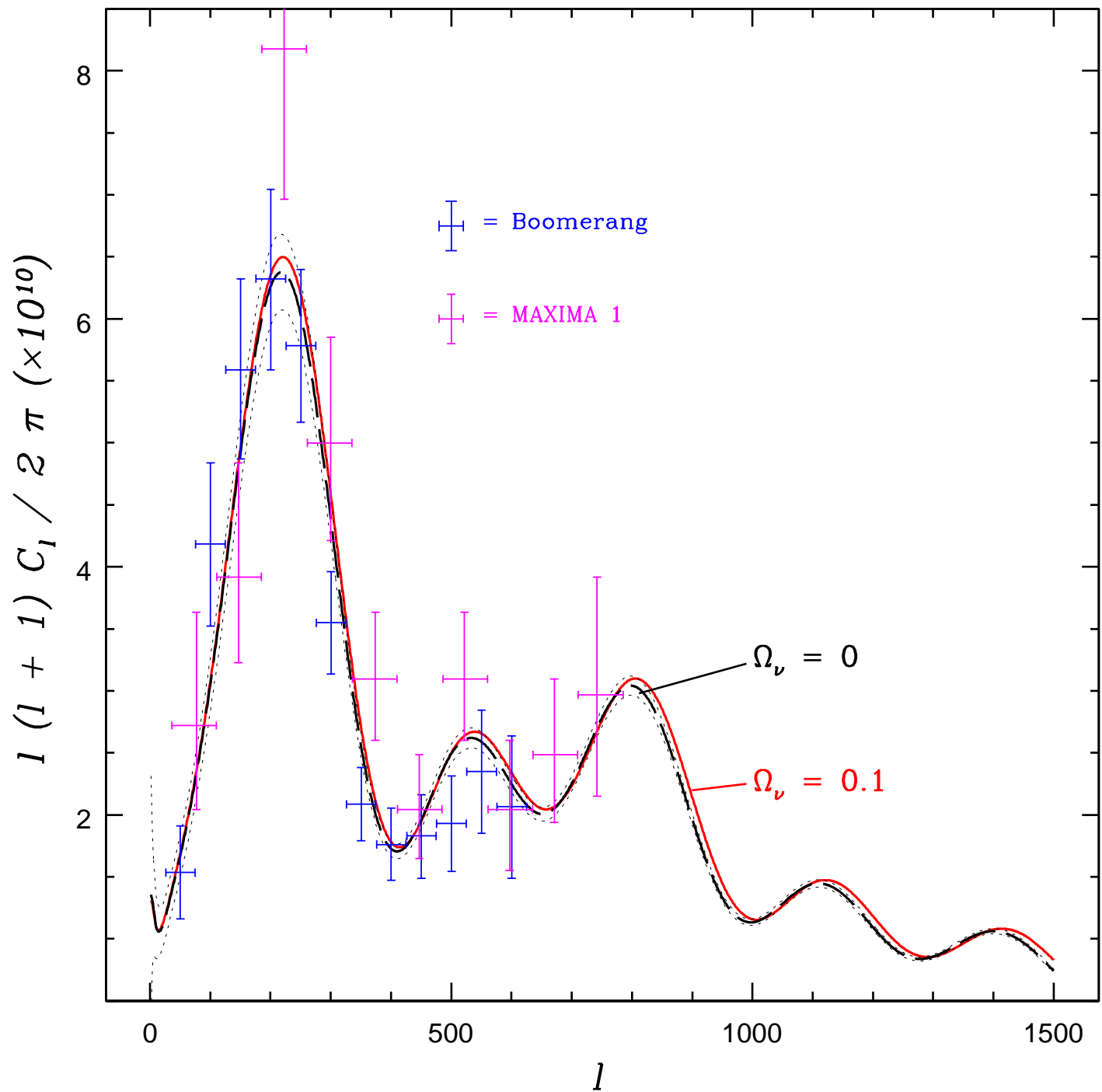}
\bigskip
\bigskip
\bigskip
\bigskip
\bigskip
\bigskip
\caption{$C_{\ell}$ spectra for the target model (black,dashed) and a true model 
with $\Omega_\nu = 0.1$ (red, solid). The spectra differ by  $\chi^2 =2300$ over 
1500 degrees of freedom, a fairly close fit to begin with.}
\end{figure}

\begin{figure}
\psfig{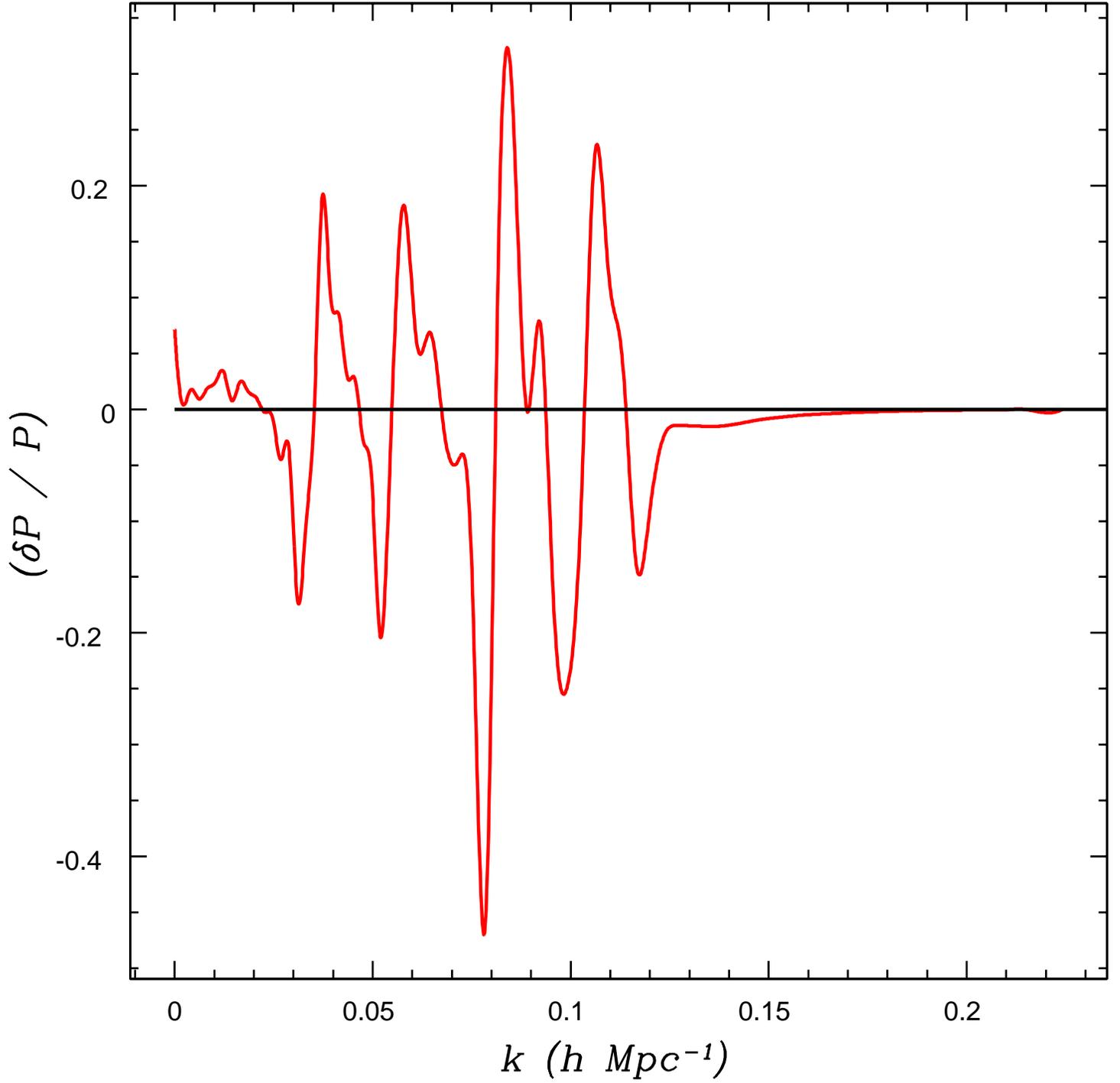}
\bigskip
\bigskip
\bigskip
\bigskip
\bigskip
\bigskip
\caption{Density power spectrum which mimics the $C_{\ell}'s$ of the target 
model ($\Omega_\nu = 0$) with parameters from the true model, $\Omega_\nu = 
0.1$. The fit has $\chi^2 = 97$ over 1500 degrees of freedom.}
\end{figure}

\begin{figure}
\psfig{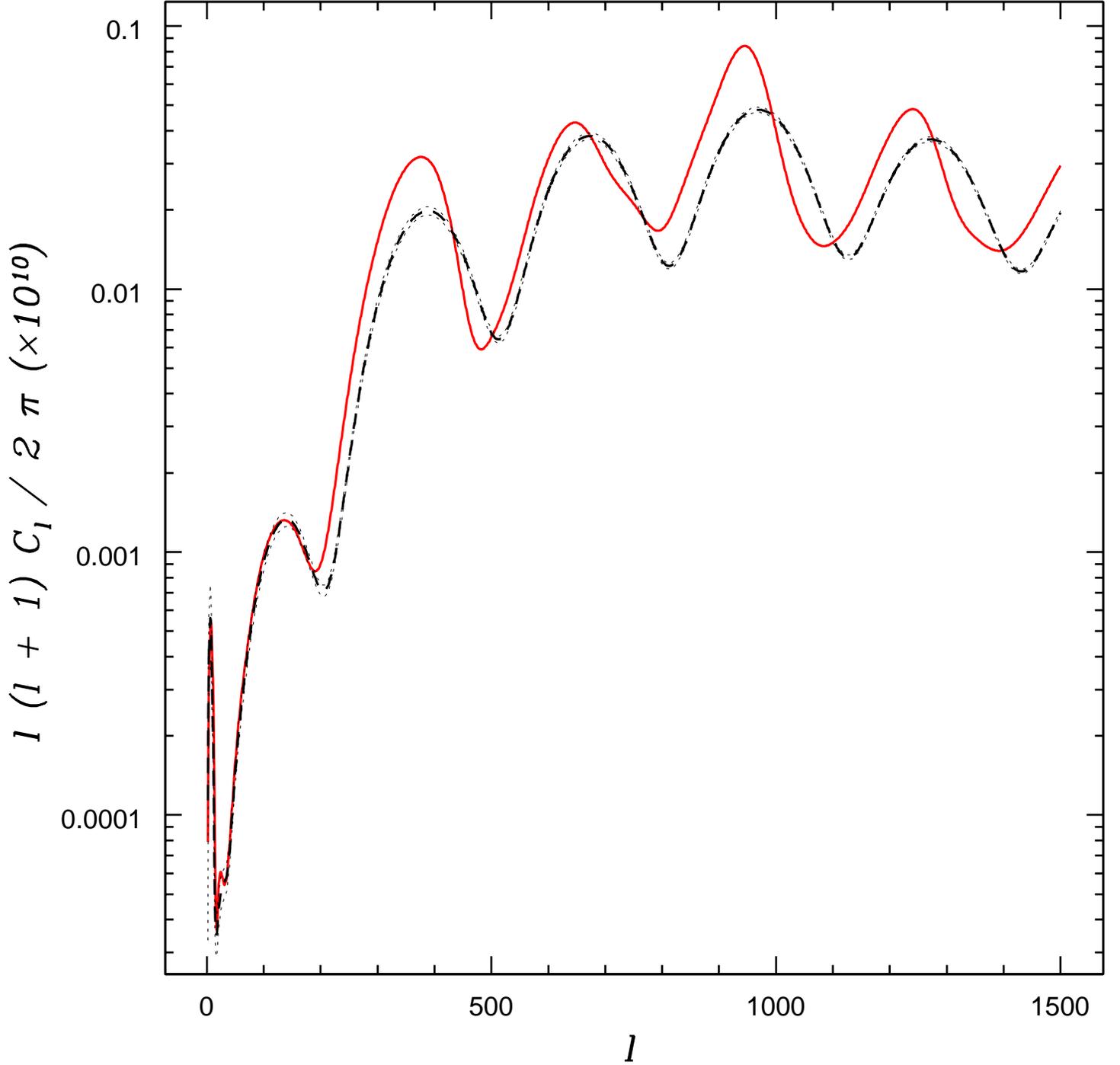}
\bigskip
\bigskip
\bigskip
\bigskip
\bigskip
\bigskip
\caption{E-mode polarization spectra for the target model ($\Omega_{\rm b} h^2 = 
0.027$, black, dashed) and the true model ($\Omega_{\rm b} h^2 = 0.019$, red, 
solid), {\it after} the power spectrum fit. The dotted lines, barely visible in 
this plot, show the cosmic variance ``envelope'' of the true model. Even though 
the temperature multipole spectra of the true and target model are 
indistinguishable, the polarization multipole spectrum differs strongly.}
\end{figure}

\begin{figure}
\psfig{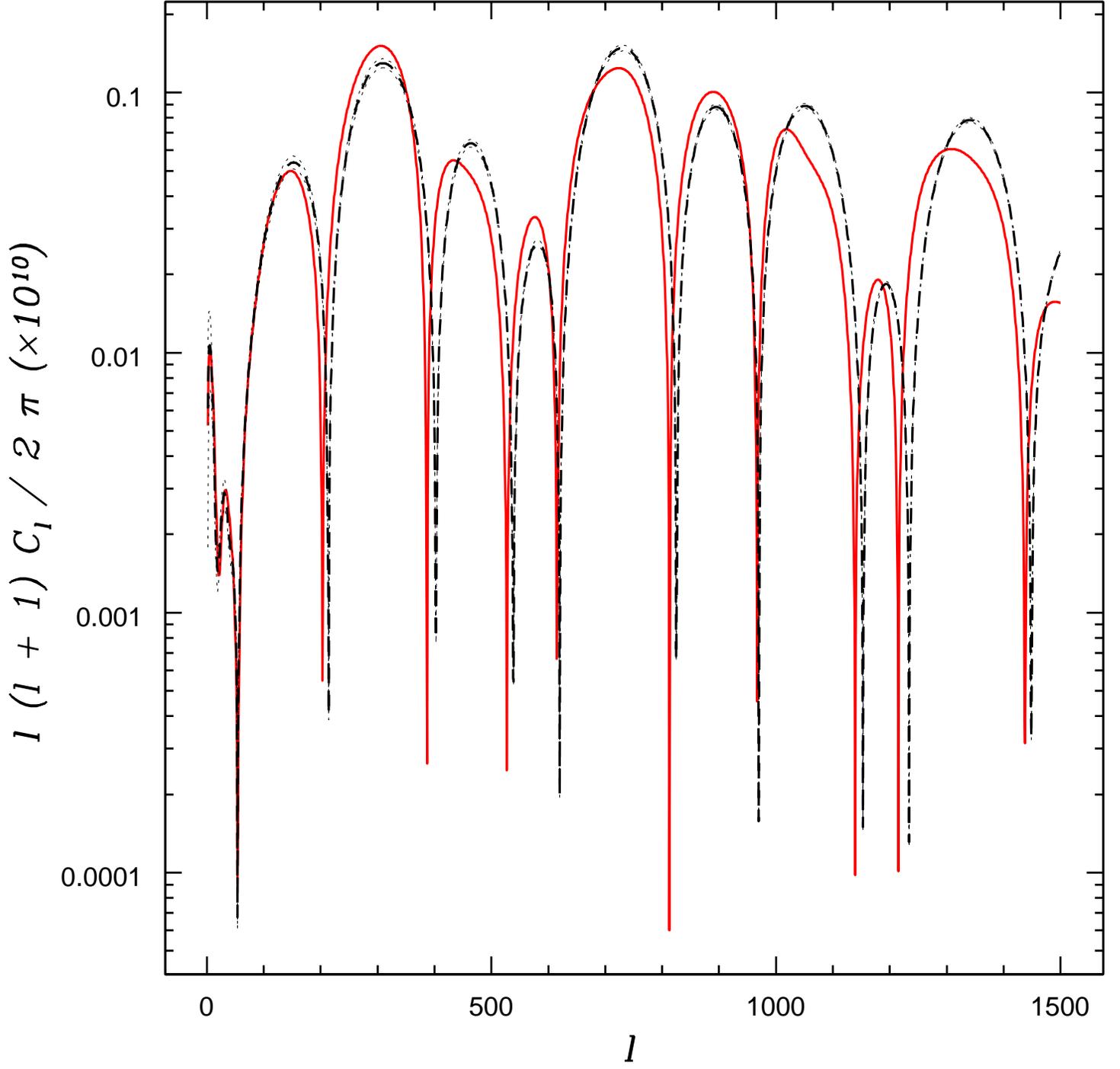}
\bigskip
\bigskip
\bigskip
\bigskip
\bigskip
\bigskip
\caption{Cross-correlation polarization spectra for the target model (black, 
dashed) and the true model ($\Omega_{\rm b} h^2 = 0.019$, red, solid), {\it 
after} the power spectrum fit, with dotted lines indicating cosmic variance 
errors. As with the E-mode polarization spectrum, the two curves are a poor 
fit.}
\end{figure}


\begin{references}
\bibitem{boom2000} P. de Bernardis {\it et al.}, Nature {\bf 404}, 955 (2000).
\bibitem{MAXIMAa} S. Hanany {\it et al.}, astro-ph/0005123.
\bibitem{boom2000a} A. E. Lange {\it et al.}, astro-ph/0005004.
\bibitem{MAXIMAb} A. Balbi {\it et al.}, astro-ph/0005124.
\bibitem{MAP} http://map.gsfc.nasa.gov/
\bibitem{Planck} http://astro.estec.esa.nl/SA-general/Projects/Planck/
\bibitem{gawiser98} E. Gawiser and J. Silk, Science {\bf 280}, 1405 (1998), 
astro-ph/9806197.
\bibitem{sloandss} http://www.sdss.org/
\bibitem{stewart97} E. D. Stewart, Phys. Lett. B {\bf 391}, 34 (1997), 
hep-ph/9606241.
\bibitem{stewart97a} E. D. Stewart, Phys. Rev. D {\bf 56}, 2019 (1997), 
hep-ph/9703232.
\bibitem{copeland97} E. J. Copeland, I. J. Grivell, and A. R. Liddle, Mon. Not. 
Roy. Astr. Soc. {\bf 298}, 1233 (1998), astro-ph/9712028.
\bibitem{covi98} L. Covi, D. H. Lyth and L. Roszkowski, Phys. Rev. D {\bf 60}, 
023509 (1999), hep-ph/9809310.
\bibitem{covi99} L. Covi and D. H. Lyth, Phys. Rev. D {\bf 59} (1999) 063515, 
hep-ph/9809562.
\bibitem{covi00} D. H. Lyth and L. Covi, astro-ph/0002397.
\bibitem{lesgourgues99} J. Lesgourgues, S. Prunet and D. Polarski, Mon. Not. 
Roy. Astron. Soc. {\bf 303}, 45 (1999), astro-ph/9807020.
\bibitem{seljak97} U. Seljak and M. Zaldarriaga, Astrophys. J. {\bf 469}, 437 
(1996), astro-ph/9603033.
\bibitem{nrcubicspline} W. H. Press, B. P. Flannery, S. A. Teukolsky, and W. T. 
Vetterling, {\it Numerical Recipes} (Cambridge University Press, Cambridge, 
1989), Sec. 3.3, p. 86.
\bibitem{nrgradientdescent} W. H. Press, B. P. Flannery, S. A. Teukolsky, and W. 
T. Vetterling, {\it Numerical Recipes} (Cambridge University Press, Cambridge, 
1989), Sec. 14.4, p. 521.
\bibitem{nrsvdcomp} W. H. Press, B. P. Flannery, S. A. Teukolsky, and W. T. 
Vetterling, {\it Numerical Recipes}, (Cambridge University Press, Cambridge, 
1989), Sec. 2.9, p 52.
\bibitem{burles98} S. Burles and D. Tytler,  Proceedings of the Second Oak Ridge 
Symposium on Atomic and Nuclear Astrophysics, (Oak Ridge, TN, December 2-6, 
1997), ed. A. Mezzacappa (Institute of Physics, Bristol), astro-ph/9803071.
\bibitem{burles99} S. Burles, K. M. Nollett, J. N. Truran and M. S. Turner, 
Phys. Rev. Lett. {\bf 82}, 4176 (1999), astro-ph/9901157.
\bibitem{kamionkowski96} M. Kamionkowski, A. Kosowsky, and A. Stebbins, Phys. 
Rev. D {\bf 55}, 7368 (1997), astro-ph/9611125.
\bibitem{zaldarriaga98} M. Zaldarriaga and U. Seljak, Phys. Rev. D {\bf 58} 
(1998) 023003, astro-ph/9803150.
\bibitem{wang98} Y. Wang, D. N. Spergel, and M. A. Strauss, Astrophys. J. {\bf 
510}, 20 (1998), astro-ph/9802231.

\end{references}
\end{document}